\documentclass[12pt,preprint]{aastex}
\usepackage{color}

\shortauthors{Zelaya {\it et~al.\/}}
\shorttitle{Spectropolarimetry of SN 2007sr}

\newcommand{\kms}{\mbox{km s$^{-1}$}}

\begin{document}

\title{Spectropolarimetry of the Type Ia SN 2007sr Two Months After Maximum Light\altaffilmark{*}}
\author{
P. Zelaya\altaffilmark{1}, 
J.R. Quinn\altaffilmark{1}, 
D. Baade\altaffilmark{2}, 
A. Clocchiatti\altaffilmark{1}, 
P. H{\"o}flich\altaffilmark{3}, 
J. Maund\altaffilmark{4},
F. Patat\altaffilmark{2}, 
L. Wang\altaffilmark{5}, and
J.C. Wheeler\altaffilmark{6} 
}

\authoremail{pazelaya@astro.puc.cl}

\altaffiltext{*}{Based on observations obtained with the VLT at the European Southern Observatory (ESO), Chile [ESO programme 080.D-0107A]}

\altaffiltext{1}{Departmento de Astronom\'ia y Astrof\'isica, Pontificia
Universidad Cat\'olica de Chile, Casilla 306, Santiago 22, Chile}

\altaffiltext{2}{ESO - European Organization for Astronomical Research in the Southern Hemisphere, Karl-Schwarzschild-Str.2, 85748 Garching b. M\"unchen, Germany}

\altaffiltext{3}{Department of Physics, Florida State University, Tallahassee, Florida 32306-4350,USA}

\altaffiltext{4}{
Queen's University Belfast, University Road, Belfast, BT7 1NN, UK}

\altaffiltext{5}{Department of Physics, Texas A\&M University, College Station, Texas 77843-4242, USA}

\altaffiltext{6}{Department of Astronomy and McDonald Observatory, The University of Texas at Austin, 1 University Station, C1400, Austin, Texas 78712, USA}

%\date{\today} 
%\maketitle

\begin{abstract}

We present late time spectropolarimetric observations of SN 2007sr, obtained with the VLT telescope at ESO Paranal Observatory when the object was 63 days after maximum light.
The late time spectrum displays strong line polarization in the Ca~II absorption features.

SN 2007sr adds to the case of some normal Type Ia SNe that show high line polarization or repolarization at late times, a fact that might be connected with the presence of high velocity features at early times.

\end{abstract}

%Section heading
\section{Introduction}

Type Ia Supernovae (SNe) are believed to form in binary systems. 
Traditional scenarios involve a carbon-oxygen white dwarfs accreting material from a companion, which could be a main sequence, a red giant, a helium, or another white dwarf star. 
As the degenerate star accretes material, it approaches the Chandrasekhar mass limit.
Once arrived at that point it becomes pressure unstable and a thermonuclear runaway starts at one or more ignition points.
A thermonuclear burning front propagates synthesizing new elements, at first subsonically in a deflagration regime, then it probably experiences a transition to a supersonic regime, a detonation \citep{Khokhlov1991A&A...245L..25K,Hillebrandt2000AIPC..522...53H}.
The deflagration phase leaves a residual of intermediate mass elements. In the
detonation, the thermonuclear burning continues until nuclear statistical equilibrium is reached. The thermonuclear burning of a white dwarf is a complex process and the details are still subject to debate.

In the last decades Type Ia SNe have successfully proved to be reliable distance indicators when standardized by their luminosity decline rate \citep{Phillips1993ApJ...413L.105P,Hamuy1996AJ....112.2408H,Prieto2005ASPC..339...69P}. 
They represent a rather homogeneous class, in particular, the subtype defined by \citet{Branch1993AJ....106.2383B} as ``spectroscopically normal'' Type Ia SNe.
Some diversity still remains in their peak luminosity, color and light curve shape, and it is attributed to different intrinsic SN properties such as progenitors, host galaxy, environment and explosion mechanism \citep{Ivanov2000ApJ...542..588I, Hamuy2000AJ....120.1479H, Hoeflich_Khokhlov1996ApJ...457..500H}.

The natural assumption of spherical symmetry, which is implicit in the use of SNe as distance estimators,
can be tested by using different tools, such as:
1) powerful hydrodynamical simulations;
2) nebular phase spectroscopy, when the ejecta becomes transparent, the continuum intensity decreases and emission lines from the deep interior of the exploded star emerge;
and 3) high quality spectropolarimetric observations collected through the last decade.
Starting from a spherically symmetric white dwarf,
hydrodynamical models produce asymmetries by off-center ignition and highly
turbulent nuclear burning during the deflagration phase \citet{Kasen2008JPhCS.125a2007K,Kasen2009Natur.460..869K}.
Coupled with viewing angle effects these contribute to observational diversity.
%{\color{red} dimmer SNe would be more asymmetric due to more burning in the turbulent deflagration phase}
Nebular phase spectroscopy, starting at about 100 days after maximum light, reveals the inner core of the supernova with signatures of asymmetries formed in these stages \citep{Maund2010ApJ725L.167M,Maeda2010ApJ...708.1703M}.
%and they show what?
%they show that blueshift of some lines can be explained by a kinematical model, correlating the viewing angle with the direction of an offset of the ignition point
%
HST imaging of the supernova remnant SN1885, S-Andromeda, revealed asymmetries in Ca and Fe which can be understood in the framework of off-center delayed detonations \citep{Fesen2007ApJ...658..396F}.
%Ca is formed in the region of incomplete Si-burning and in the low temperature end of NSE which produces 56Ni). Therefore, it is be mostly sensitive to off-center DDTs.
%
Spectropolarimetric observations show large diversity, except maybe in the continuum polarization, where low values are common.
According to theoretical interpretation
\citep{Hoflich1991A&A...246..481H,Howell2001ApJ...556..302H},
This indicates that the photosphere is not completely spherical, and
small global asymmetries, less than 15\%, are inferred.
A contrasting picture is revealed by the polarization in the blueshifted absorption troughs of the P Cygni profiles where a large diversity is observed.
This is usually interpreted as the result of small scale departures of
symmetry, chemical or excitation inhomogeneities. 
Spectropolarimetric data show line polarization for different elements, such as
Si~II, Ca~II, S~II, with different values at different epochs of the supernova
evolution. 
The general consensus emerging from the data is that
Type Ia SNe are more asymmetric in their outer layers, and more spherically symmetric in the interior.
This implies that line and continuum polarization are higher at early times and decrease towards later epochs when 
the photosphere recedes into the inner layers 
\citep{Filippenko_Leonard2004cetd.conf...30F,Leonard2005ApJ...632..450L,Wang_and_Wheeler2008ARA&A, Wang2003ApJ...591.1110W}.
However, SN 2006X, one of the few Type Ia SNe that has been observed in spectropolarimetry at times later than 30 days after maximum, does show strong
late-time line polarization in the Ca~II IR triplet \citep{Patat et2009A&A...508..229P}.

In this paper, we report spectropolarimetric observations of SN 2007sr
when the SN was more than two months after maximum and showed highly
polarized Ca~II lines.
In Section \ref{sec:Data} we present the data, in Section \ref{sec:Results} we describe the results, and in Section \ref{Sec:Discussion} we discuss them and
present our conclusions.

\section{Data}\label{sec:Data}

\subsection{Supernova 2007sr}
%Aspects of 07sr

Supernova 2007sr was discovered in the Antennae galaxies (NGC 4038/39) on December 18.53 UT by the Catalina Sky Survey
\citep{Drake2007CBET.1172....1D}.
It was classified as a type Ia about 3--4 days after $B$ maximum light
\citep{Pojmanski2008CBET.1213....1P,Umbriaco2007CBET.1174....1U}. It spectroscopically matched normal Type Ia SNe with a Si~II velocity gradient $\dot{v}_{\rm{SiII}}=80\pm15$ \kms day$^{-1}$ \citep{Maeda2010Natur.466...82M}, which following \citet{Benetti2005ApJ...623.1011B} places it among the High Velocity Gradient (HVG) events.
\citet{Schweizer2008AJ} calibrate the intrinsic brightness of SN~2007sr using the analysis package SNooPy \citep{Burns2011AJ....141...19B}, and following \citet{Folatelli2010AJ....139..120F}, assume a low value for the ratio of total to selective extinction $R_{V}^{\rm{host}}$ = 1.6 $\pm$  0.1 .
They find a decline-rate parameter $\Delta m_{15}$ = 0.97 mag $\pm$ 0.02 mag, and a color excess in the host galaxy $E(B-V)_{host}$ = 0.13 mag $\pm$ 0.01 mag.
The color excess caused by our own galaxy is $E(B-V)_{MW}$ = 0.046 $\pm$ 0.005, according to \citet{Schlegel1998ApJ...500..525S}.

The distance to SN~2007sr found by \citet{Schweizer2008AJ} is D = 22.3 $\pm$ 2.8 Mpc.
This makes the host consistent with the large-scale flow model of \citet{Tonry2001ApJ...546..681T}, in
agreement with four other nearby galaxies that, according to their recession velocities, belong into the same
group \citep{Madore_and_Steer2007}.
Some discrepancy was found with the result of \citet{Saviane2008ApJ...678..179S} who estimated a much smaller distance
using the tip of the red giant branch, D = 13.3 $\pm$ 1.0 Mpc. 
% CHECK HOY
The difference in the distance estimates has no effects on the results presented here.

\subsection{Spectropolarimetry of SN 2007sr}
Spectropolarimetric observations of SN 2007sr were obtained with the ESO VLT UT1 telescope, on 2008 February 15
using the FORS1 instrument in polarimetry mode\footnote{For details of the instrument check the FORS Manual at\\ http://www.eso.org/sci/facilities/paranal/instruments/fors/doc/}.
According to the light curve measured by the Carnegie Supernova Project \citep[Table 1 in][]{Schweizer2008AJ}
our observations took place when the supernova was 63 days after $B$ maximum light, which corresponds to the spectroscopic Fe~II phase \citep{Branch2005PASP..117..545B,Branch2008PASP..120..135B}. At this epoch the spectrum of Type Ia's is mainly shaped by Fe~II, Ca~II and Na~II lines in addition to forbidden lines.

Four exposures of $\sim$ 1200 s were taken with the retarder plate at position angles of 0, 22.5, 45 and 67.5 degrees.
Grism GRIS\_300V was used with no order separation filter.
This provides a dispersion of 3.28 \AA{}/pix, and the resulting wavelength range 350--900 nm.
The spectral resolution with a 1$''$ slit is 12.5 \AA{}, which we measured using the OI~5577 sky emission line.
Additional observations of a flux standard, GD 108, were performed at position angle 0 degrees to calibrate the flux level.

%http://www.eso.org/sci/facilities/paranal/instruments/fors/inst/grisms.html

We followed a standard data reduction procedure.
Bias, prescan and overscan corrections, flat fielding, spectrum extraction and wavelength calibration were performed using IRAF\footnote{ http://iraf.noao.edu---IRAF is distributed by the National Optical Astronomy Observatories,which are operated by the Association of Universities for Research in Astronomy, Inc., under cooperative agreement with the National Science Foundation.}. The sensitivity curve of the spectrograph was computed from the flux standard observed at one position angle.

Polarized spectra and flux calibration were obtained with our own IDL routines. Stokes parameters were calculated as described in
\citet{PatatRomaniello2006PASP..118..146P}.
As a starting point a bin size has to be selected. This is a critical step. When polarization is present it should be detected using any bin size. But the estimated polarization
values, in particular the peak values, are sensitive to the bin size.
We have chosen a bin size of 15 \AA, slightly larger than the resolution. 
The spectrum at zero position angle was binned to 15 \AA\ and the 
spectra at the other position angles were resampled to these same bins.
The resampling was done using a five-point closed Newtonian-Cotes algorithm.

Observed Stokes parameters
are calculated using the formulas given in the VLT FORS User Manual, including
the correction for the position angle zero point.
The sensitivity function is applied to each beam, together with atmospheric extinction corrections for Paranal as suggested by \citet{Patat2011A&A...527A..91P}.
The uncertainties of the Stokes
parameters Q and U were calculated assuming a Gaussian distribution for the counts at each wavelength.

\section{Results}\label{sec:Results}
\subsection{Intensity Spectrum} \label{Intensity_spec}
\begin{figure}[b]
\centering
    \epsscale{0.8}
    \plotone
%{/home/lavioleta/polariz/2007sr/figures/spec_sky_signalnoiseALL.eps}
{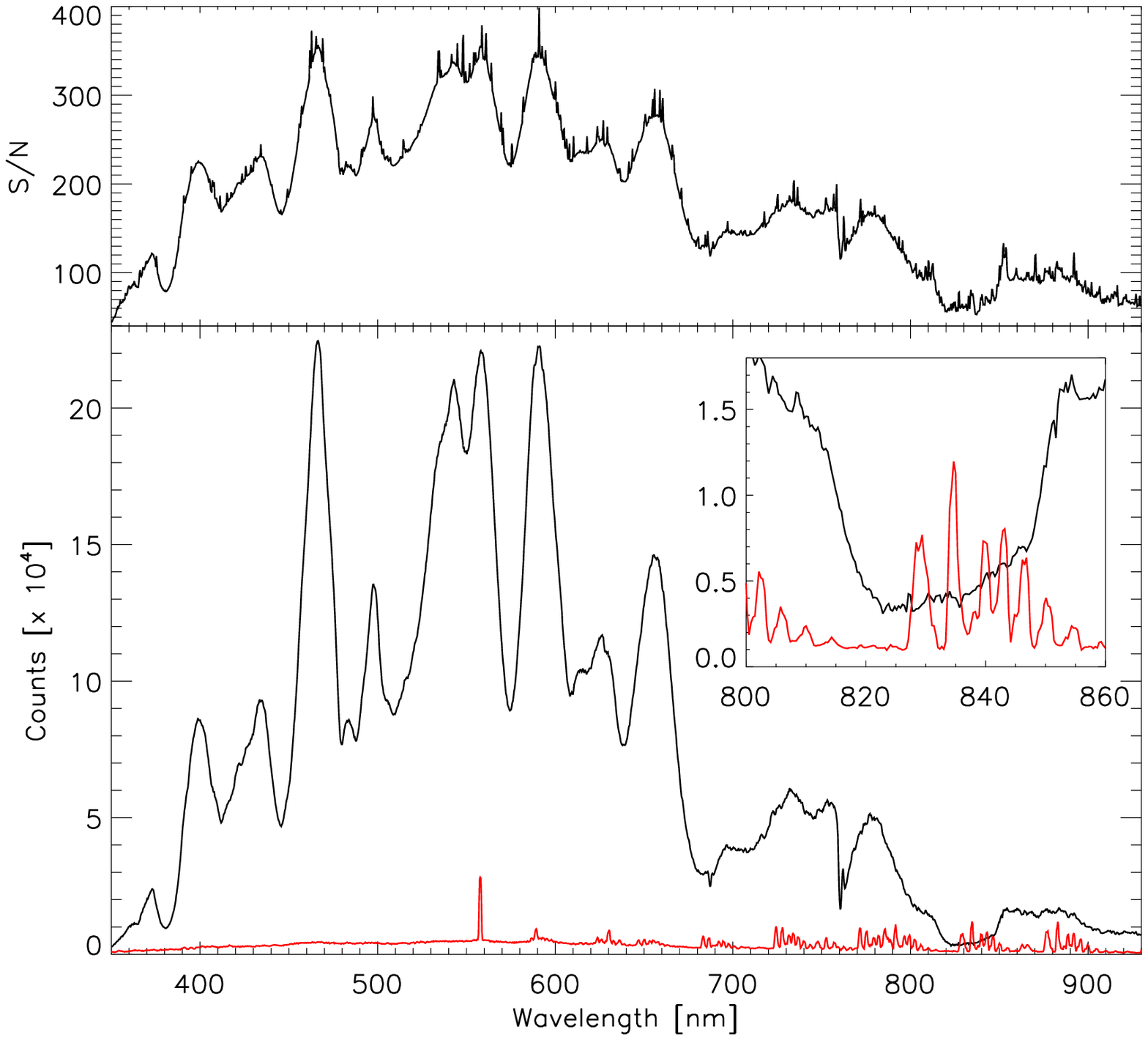}
\caption{Intensity spectrum of SN 2007sr and the sky in counts (lower panel) and the S/N ratio (upper panel). In the region of the Ca~II IR triplet (see inset in lower panel) the S/N decreases significantly and the sky emission lines contaminate the red portion of the blueshifted absorption profile.}
\label{fig:sky+spec}
\end{figure}

Figure \ref{fig:sky+spec} shows the observed intensity spectrum in counts and the signal to noise ratio (S/N).
The latter (S/N $\sim$ 100-300) is rather low for the standards of spectropolarimetry (generally S/N $>$ 500), as a result of the SN being dimmer than estimated when
observations were done ($M_B \sim 16.2$).
We note that in the red portion of the spectrum, from 700--950 nm, approximately, some of the sky emission lines approach to, or exceed, the supernova flux.
Background contamination can be an issue there.
The blueshifted absorption minimum of the Ca~II IR triplet P Cygni profile,
in particular, reaches just a few thousand counts,
and is contaminated by sky emission light in a large fraction of the redder
part.
Yet some 100 \AA{} of the bluest edge of the absorption feature contain signal above the sky with S/N~$\sim 60$.
This range, between 8150--8250 \AA{} will be considered as a reliable
zone to measure line polarization.

In Figure \ref{fig:latespecs}, we display the intensity spectrum of SN~2007sr and compare it with those of other Type Ia SNe at a similar phase.
Following line identifications from the literature \citep{Liu1997ApJ...489L.141L, Hatano1999ApJ...525..881H, Stanishev2007A&A...469..645S}
we identify lines of Si~II, Ca~II, Na~I, Fe~II, Fe~III, Co~II, Co~III and O~I at velocities around 13000--10000 \kms, as indicated in the figure. 

We detect two systems of weak and narrow Na~I D lines in absorption, unresolved by our spectrograph (see Figure \ref{fig:NaDlines}).
One of them, at observed wavelength of 589.19 nm, is consistent with interstellar matter in the Galaxy.
The other, at observed wavelength 592.55 nm, is consistent with interstellar matter in NGC~4038 and implies a recession velocity of 1705 $\pm$ 144 \kms. 
This velocity agrees with that of the NASA/IPAC Extragalactic Database, 1642 $\pm$ 12 \kms \citep{LaubertsandValentijn1989Msngr..56...31L} .

%z=(5892.-5925.5)/5892.*299792.458 
%dz= sqrt(((5925.5/5892^2)*5)^2+((1/5892.)*4)^2)  dlambda1=5AA dlambda2=4 AA

%NaID SN (11000 kms) center: 5744.5   flux: -8.1E-13 +- 0.3E-16 eqw: 102.5  +-0.9   gfwhm: 156.4 +- 1.5
%     MW             center: 5892.25  flux: -9.5E-16 +- 1E-16   eqw: 0.1206 +-0.02  gfwhm: 11.88  +- 1
%     Host           center: 5925.2   flux: -9.6E-16 +- 1E-16   eqw: 0.1229 +-0.02  gfwhm:  8.665 +- 1
%Turatto2002
%E(B-V)= -0.04 + 0.51EW(NaID)= -0.04+0.51*(0.1229) = 0.0226790   (9x() = 0.204111 )
%E(B-V)= -0.01 + 0.16EW(NaID)= -0.01+0.16*(0.1229) = 0.00966400  (9x() = 0.086976 )
%Schweizer 2008
%E(B-V)= 0.13   (9x() = 1.17 )

\subsection{Polarization}

The observed Stokes parameters, and polarization are shown in Figure \ref{fig:fluxQUPT_ep1}.
Going from top to bottom, the panels show the observed flux spectrum,
the observed polarization,
the observed Q and U Stokes parameters, and the observed polarization angle. 

Establishing the intrinsic polarization of the supernova
requires the subtraction of additional polarization sources,
like the polarization caused by interstellar matter in our galaxy
and the host (this is the so-called interestellar polarization (ISP)). 
We estimate a value for the ISP based on two assumptions.
One is that the emission part of the strong P-Cygni profiles in the spectrum
are intrinsically unpolarized.
This should be so, because they originate in resonant scattering of photons
coming from the photosphere, which are emitted with random orientation of the
plane of polarization.
This spectrum has two good lines to estimate the ISP in this way:
the blend of Na~I~D and Co~III at
rest frame wavelength $\sim$5700 \AA{} and the Ca~II H\&K.
The other assumption is that the blue side of the spectrum, the region between
$\sim$400 and $\sim$650 nm, approximately, includes
the blends of P-Cygni profiles of many different lines.
Since blending of P Cygni lines should have a net depolarizing effect due to line
blanketing opacity \citep{Maund2010ApJ...722.1162M} it is reasonable to assume zero
intrinsic polarization in that part of the spectrum.
Both assumptions consistently indicate an ISP with Stokes parameters Q$_{ISP} \simeq 0$ and U$_{ISP} \simeq 0.1$ (P$_{ISP}$=0.1).
Finally, we assume a single Serkowski profile with $\lambda_{\rm max}$  = 5500 \AA{} as the wavelength dependence of the ISP \citep{Serkowski1975ApJ...196..261S}.
Panel 2 of Figure \ref{fig:fluxQUPT_ep1} shows the ISP (nearly horizontal red line), which confirms that the intrinsic polarization of the continuum is, indeed, consistent with zero.

We note that the previous ISP estimate is consistent with the observational relation between color excess $E(B-V)$ and P$_{ISP}$
established by \citet{Serkowski1975ApJ...196..261S}, which takes the form of the upper limit $P_{ISP, max}/E(B-V) \lesssim 9.0$.
Both the color excess in the SN host measured by \citet{Schweizer2008AJ}, and the one in the Galaxy, could justify an even larger ISP.
We note, as well, that the equivalent width of the narrow Na~I~D lines in our spectrum, and the color excess of Schweizer et al.,
give conditions for SN~2007sr for which both correlations in \citet{Turatto2003fthp.conf..200T} are appropriate (see their Fig. 3).
Turatto's relations, which can be used to calculate two extreme values of E$(B-V)$,
give the color excess E$(B-V)$ expected for a given EW of the
NaI D lines.
The two linear correlations they find are probably the result of to different dust-to-gas ratios in the host galaxies.

In contrast with the null continuum polarization, there is line polarization,
in the Ca~II features.
With the spectrum binned to 15 \AA,
the Ca~II H\&K and the infrared triplet display polarized components
with maximum values of $\sim1$\%, and $\sim4$\%, respectively.

The polarization level is reduced by a factor of two, approximately,
for a binsize of 30 \AA, see also Figures \ref{fig:CaIITripletlines}
and \ref{fig:CaIIHKlines}.
In any case, the polarization angle is close to 140 degrees. 
As pointed out in Section \ref{Intensity_spec}
the Ca~II lines extend between 5000 and 20000 \kms.
In the case of the Ca~II IR triplet, the low S/N means a noisy signal. In the
red part of the absorption profile, which corresponds to low expansion velocities,
this could even result in some residual contamination from the,
also highly polarized, sky background (see below).
The high velocity portion of the blueshifted absorption minimum, on the other hand,
is cleaner.
Finally, the Ca~II H\&K lines do not have significant background contamination.

To understand the effects of the sky in the estimated polarization, we study
the intrinsic polarization versus intrinsic polarization angle at different wavelengths, as shown in Figure \ref{fig:ThetavsPol} for SN 2007sr,
and in Figure \ref{fig:skyThetavsPol} for the sky. 
In Figure \ref{fig:ThetavsPol} we identify two well-defined angles with
significant polarization, the first around 140 degrees and the second around
45 degrees. 
The region of 140 $\pm$ 30 degrees, spanning most of the features of interest,
has been colored in gray.
The red wavelength region, 800--890 nm, which includes the Ca~II IR triplet
(red dots) displays components with significant polarization at both angles.
The blue wavelength region of the spectrum displays components with
significant polarization only around 140 degrees (bluer dots with $\sim 1$\%
polarization).
As Figure \ref{fig:skyThetavsPol} shows, the background sky that we subtracted
from the spectra to get the intensity with each position of the half wave
plate, is also significantly polarized, with an angle close to 45 degrees.
In some parts of the red portion of the Ca~II IR triplet the sky is brighter
than the supernova (see inset in Fig.~\ref{fig:sky+spec}) it is not
surprising that, there, the
observed polarization is very noisy and tends to recover the value of the sky.
This is why we prefer not to put any weight into the polarization observed
in the redder part of the Ca~II IR triplet (red points at lower angles
in Figure \ref{fig:ThetavsPol}, outside of the gray region.
We suspect that this conspicuous polarization feature
is just residual contamination of the, brighter,
subtracted sky, which happens to be polarized at about the same angle.

{
Recent work by \citet{Tanaka2012ApJ...754...63T} showed that,
for a sample of stripped-envelope SNe, stronger lines exhibited higher polarization.
They propose a simple radiative transfer model to relate the measured
polarization to more physically meaningful parameters: an enhancement
of absorption characterized by a factor $f$, and the fraction $\Delta S$, of
the photospheric disk area $S$, covered by this enhancement.
A combination of these two parameters is incorporated into
a ``corrected'' line polarization, P$_{\rm corr}$, which encodes
the asymmetry that originates the observed line polarization.
The latter, in addition, strongly depends on the strength of the
absorption feature in the intensity spectrum, measured by the fractional
depth with respect to the local continuum.
This is interesting, since it makes it possible to understand a
diversity of polarization signals resorting to a single anisotropy
and a variety of absorption line strengths.

Although the model of \citet{Tanaka2012ApJ...754...63T} was imagined and
tested for core-collapse SNe, it is general enough to be readily
applied to Type Ia ones.
We have done so, and the results of the exercise are shown in Figure~\ref{fig:TanakaPcorr}.
Measuring the fractional depth requires estimating the continuum at the
position of the absorption feature.
For the complex late spectrum of a Type Ia SN,
this is difficult to do, 
without a detailed radiative transfer model.
This is especially true in the case of the Ca~II IR triplet, where
the uncertainty in identifying the continuum leads to a fairly large
uncertainty in the fractional depth.
Even though, it is possible to see in
Fig.~\ref{fig:TanakaPcorr} that the polarization signal observed in
both Ca~II lines are mutually consistent.
Given the large difference in fractional depths,
the $\sim$1\% polarization observed for Ca~II H\&K, translates into a
2-4\% polarization in the Ca~II IR triplet,
for a modest value of the intrinsic polarization.
This reinforces the view that Ca~II is intrinsically polarized, and that
a single anisotropy is responsible of both signals.
}

The intrinsic Q-U diagram of the entire spectrum is shown in Figure \ref{fig:QUdiagram}. The scatter of points around (Q,U) =(0,0) reveals an almost null global polarization. There is a peculiar quasi-spiral loop in the region of the Ca~IR triplet, which, as mentioned earlier, we understand as due to the mixture
of intrinsic Ca~II IR triplet
polarization and sky polarization. As in Fig.~\ref{fig:ThetavsPol} we have colored gray the region where the intrinsic line polarization of the SN is located.
Figure \ref{fig:QUdiagramCaIR} gives the QU diagram of the region of Ca~II in velocity space, the IR triplet region is represented by filled circles, and the H\&K region by stars. Cyan and blue points correspond to velocities higher than 13000 \kms, uncontaminated by the sky in the case of IR triplet. It is precisely at those velocities where the strong polarization signals are measured, and seem to be concentrated in the gray area (between 110 and 170 degrees) for both Ca~II lines.

\section{Discussion and Conclusion}\label{Sec:Discussion}

\begin{figure}[b]
\centering
     \epsscale{0.55}
     \plotone
%{/home/lavioleta/polariz/2007sr/figures/latespecs.ps}
{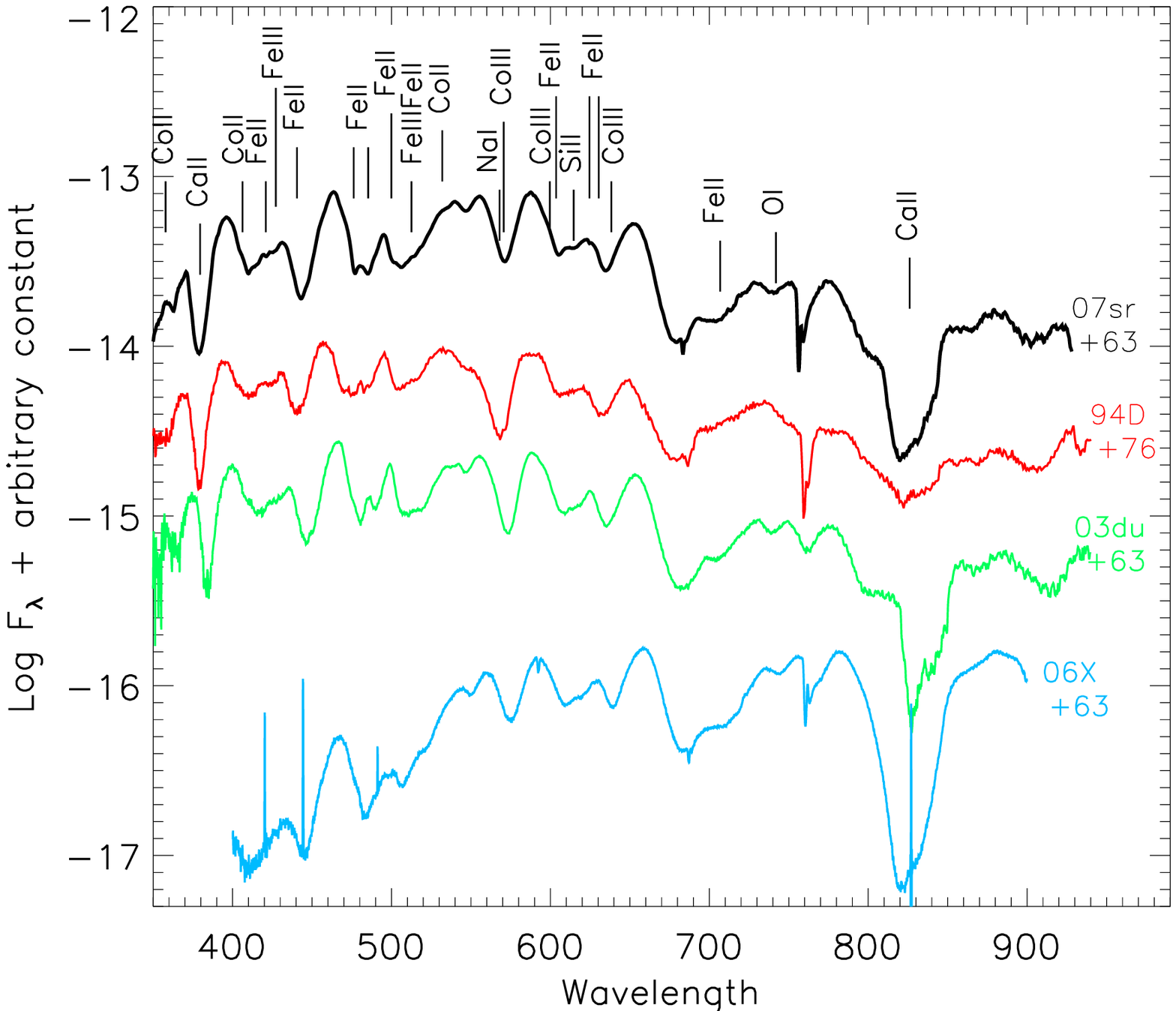}
\caption{Comparison of the flux calibrated spectrum of SN 2007sr with those of SN 1994D, 2003du and 2006X at similar phases. The data were obtained from the Online Supernova Spectrum Archive (SUSPECT), contributed by \citet{Gomez1998AJ....115.1096G,Stanishev2007A&A...469..645S,Yamanaka2009PASJ...61..713Y} respectively. All of these SNe exhibited high velocity features ( $\geq$ 20000 \,\kms) at their early phases which encourages us to infer the same behaviour for SN 2007sr (see Section \ref{Sec:Discussion}).}
% suggesting this may also be the case of SN~2007sr.}
\label{fig:latespecs}
\end{figure}

\begin{figure}[b]
\centering
     \epsscale{0.42}
     \plotone
%{/home/lavioleta/polariz/2007sr/figures/NaDlineszoom.ps}
{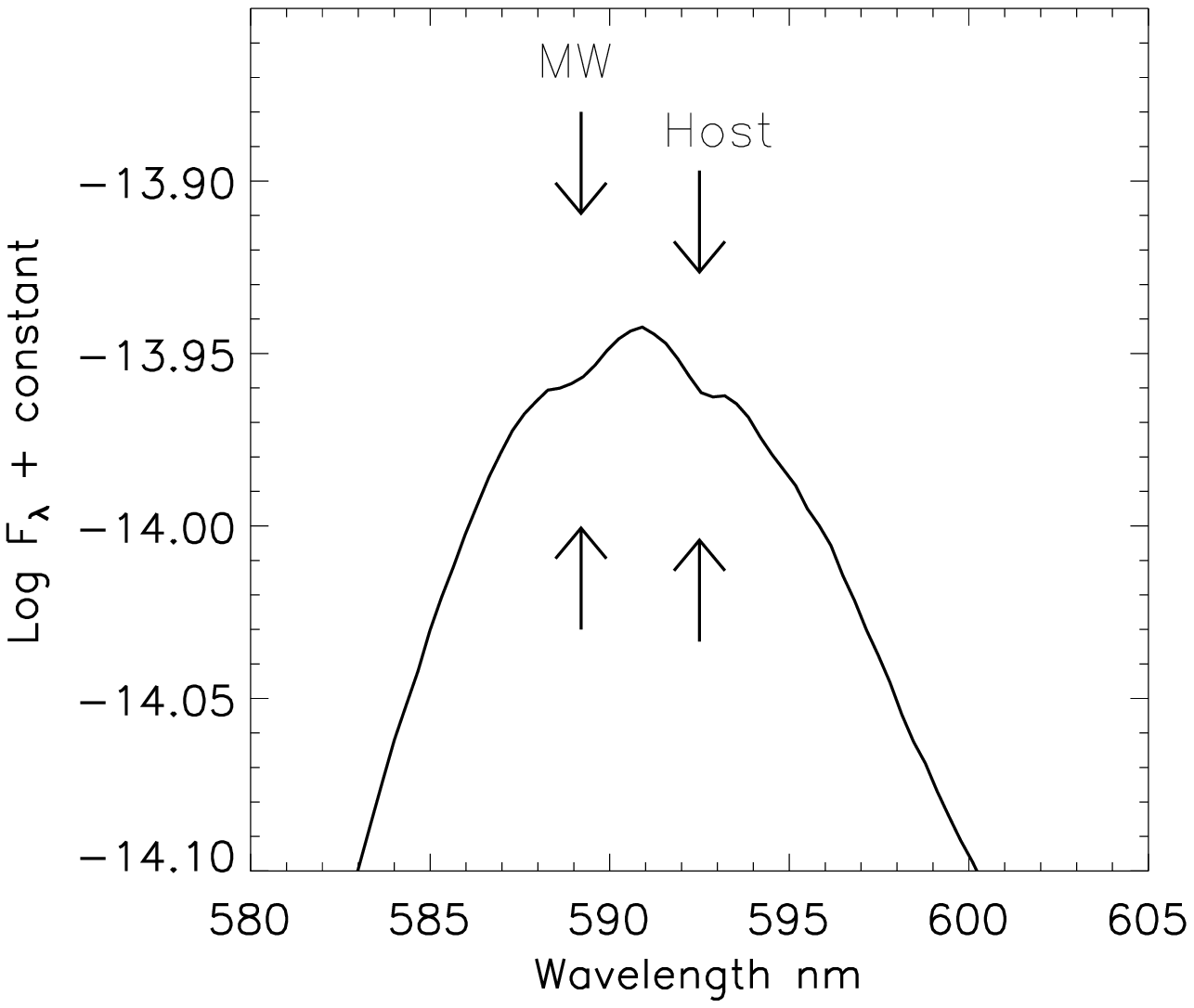}
\caption{Na~I~D lines in the Milky Way and host galaxy at 589.19 nm and  592.55 nm, respectively.
The latter implies a recession velocity of 1705 $\pm$ 144 \kms, which
agrees with the host velocity measured by \citet{LaubertsandValentijn1989Msngr..56...31L} }.
\label{fig:NaDlines}
\end{figure}

\begin{figure}[t]
\centering
    \includegraphics[height=19cm]
    %\plotone
%{/home/lavioleta/polariz/2007sr/figures/plotPolnewSN2007sr_15ep1.ps}
{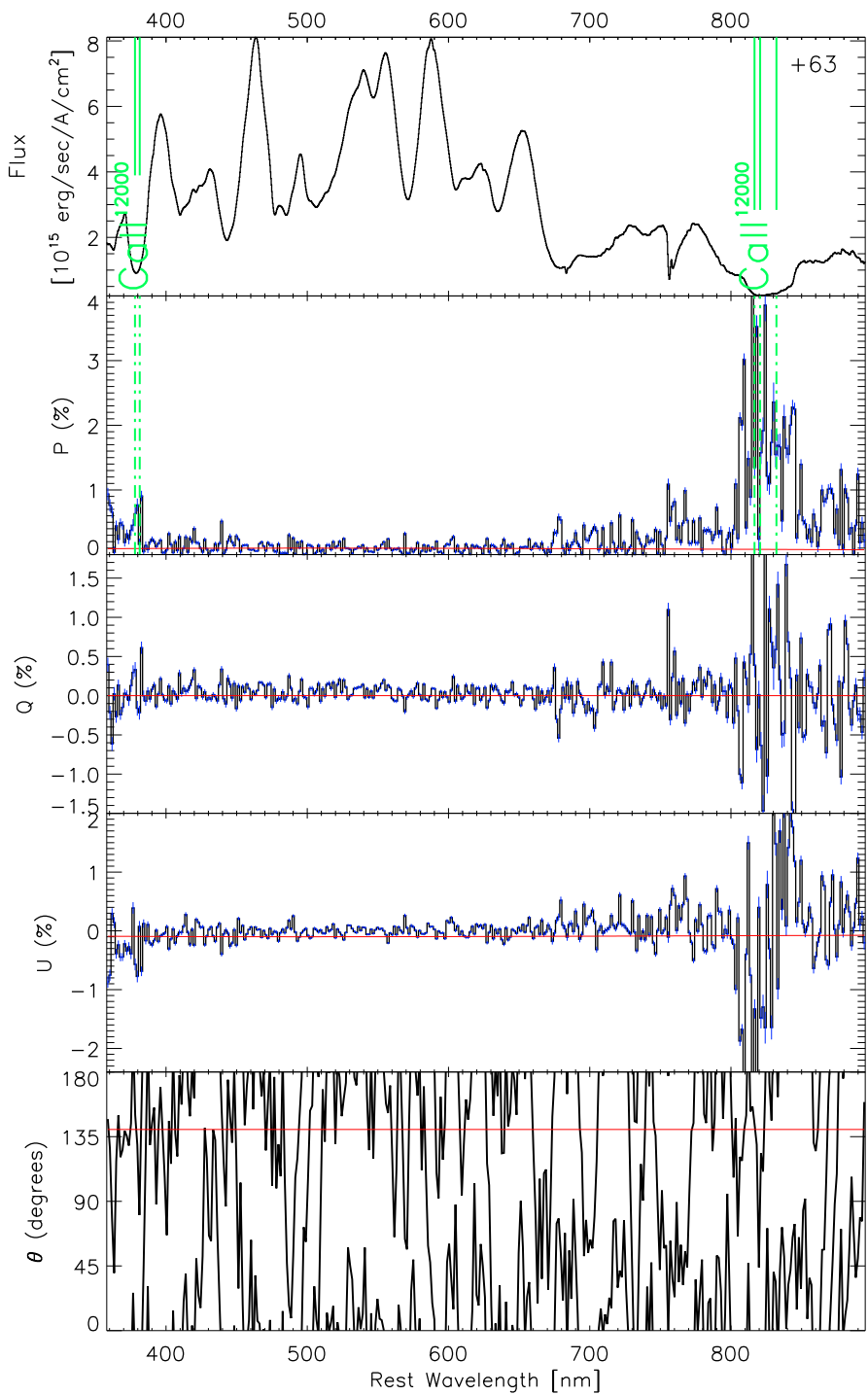}
\caption{Observed flux, polarization, Q and U parameters versus rest frame
wavelength. The Stokes parameters were computed using a bin size of 15\AA\ in
observer frame. The horizontal red line shows a Serkowski law computed with Q$_{ISP}$=0 and U$_{ISP}$=0.1. Strong polarization is observed in the region of the Ca II IR triplet, and a less prominent feature appears in the region of Ca II H\&K. Peaks around 6800 \AA{} nm and 7500 \AA{} are due to telluric absorption lines. }
\label{fig:fluxQUPT_ep1}
\end{figure}

\begin{figure}[b]
\centering
    \epsscale{1.1}
%    \plotone{/home/lavioleta/polariz/2007sr/figures/CaIIAngPOlFlux15ep1.eps}
%{f5.ps}
    \plottwo
%{/home/lavioleta/polariz/2007sr/figures/CaIIAngPOlFlux15ep1.eps}{/home/lavioleta/polariz/2007sr/figures/CaIIAngPOlFlux30ep1.eps}
{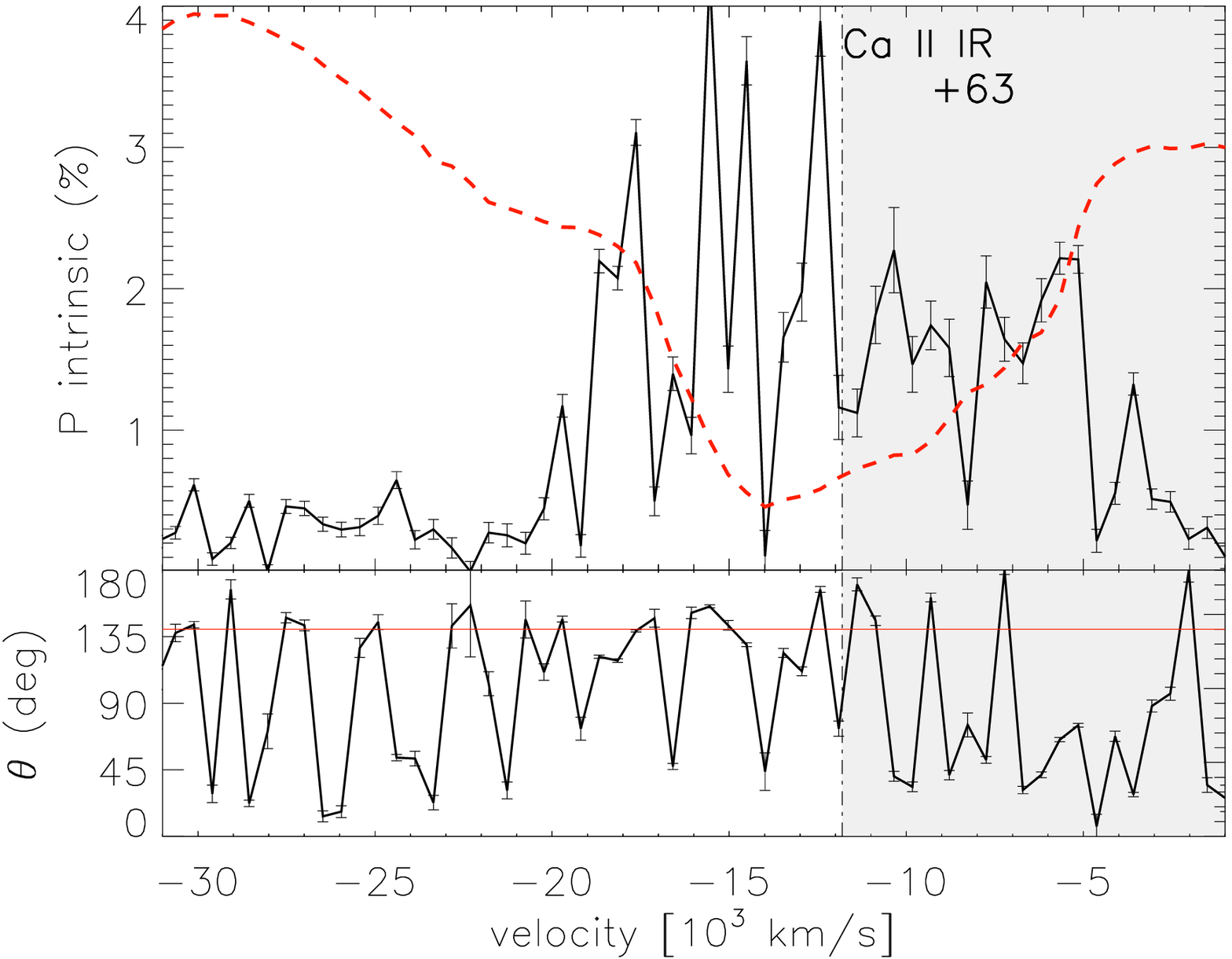}{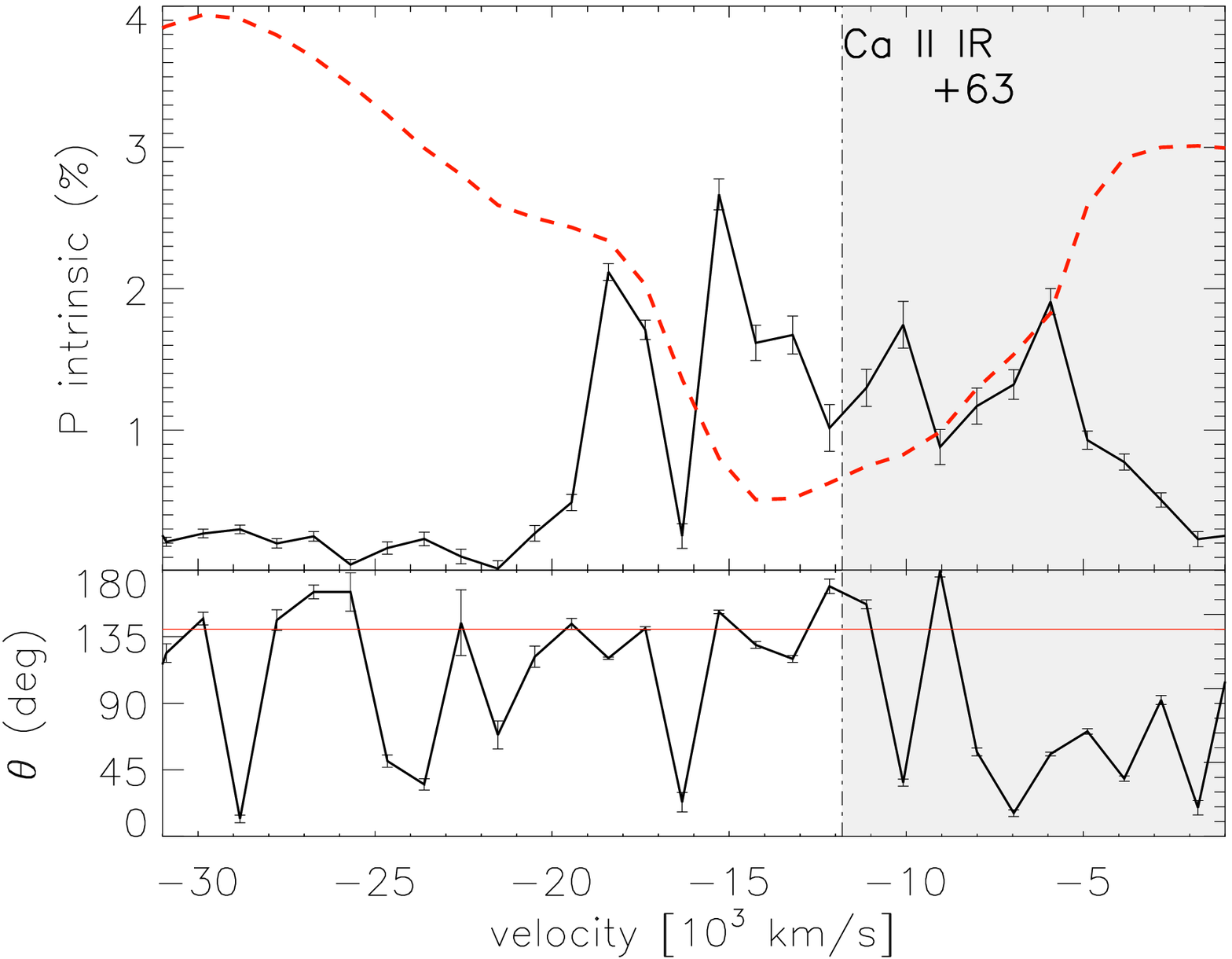}
\caption{Polarization \% and polarization angle of Ca~II NIR were computed with a bin size of 15\AA{} (left panel) and 30 \AA{} (right panel). Velocities $<$ -12000\kms\,\, are not contaminated by the sky spectrum and therefore belong to a region considered reliable for polarization measurements. The shaded area is the range of wavelength highly contaminated by sky emission lines (See Fig.~1). }
\label{fig:CaIITripletlines}
\end{figure}

\begin{figure}[b]
\centering
    \epsscale{1.1}
    \plottwo
%{/home/lavioleta/polariz/2007sr/figures/CaIIHKAngPOlFlux15ep1.eps}{/home/lavioleta/polariz/2007sr/figures/CaIIHKAngPOlFlux30ep1.eps}
{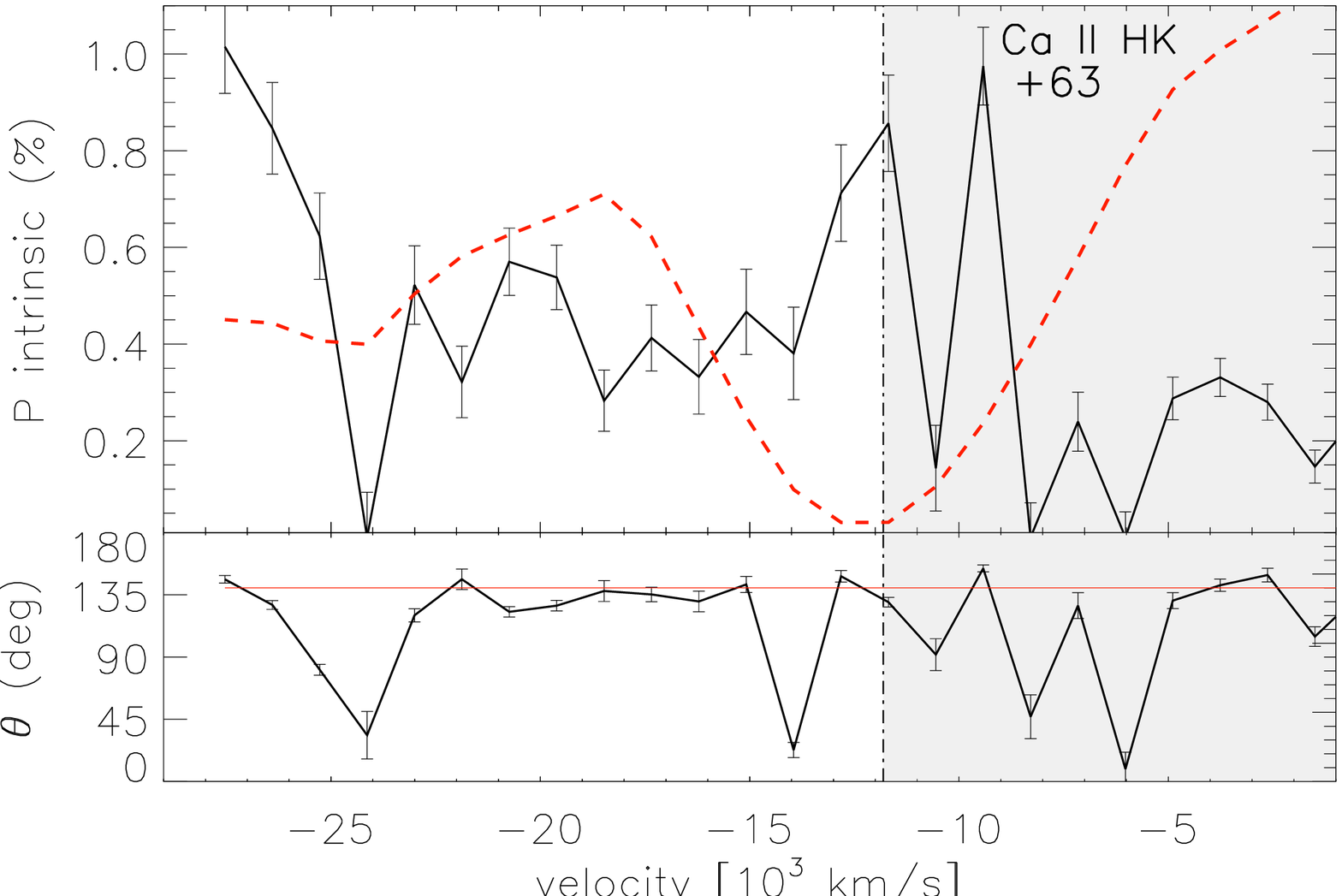}{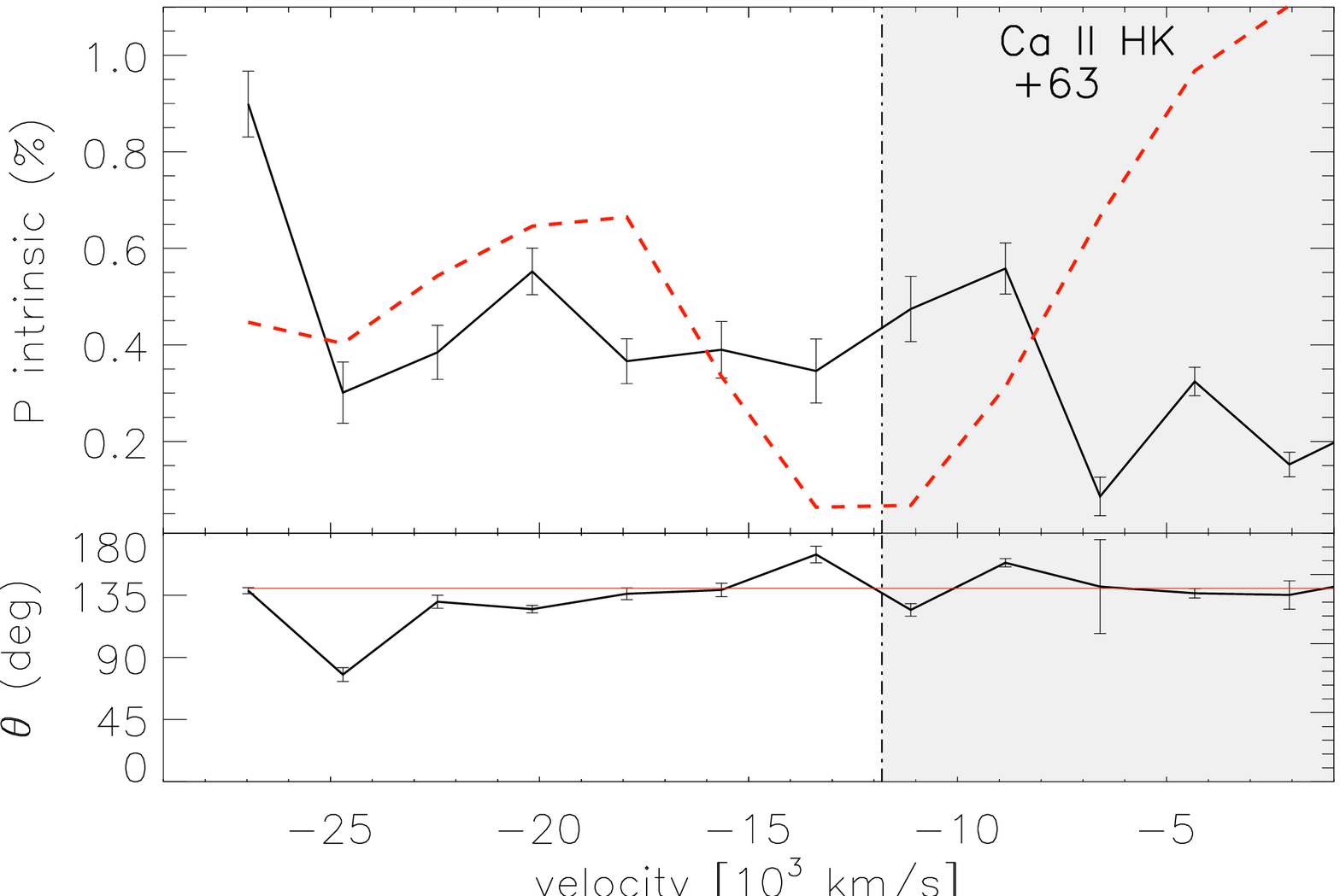}
\caption{Polarization \% and polarization angle of Ca~II H\&K were computed with a bin size of 15\AA{} (left panel) and 30 \AA{} (right panel). The polarization peaks at velocities close to -12000 \kms and the same region shown in Fig~\ref{fig:CaIITripletlines} is shaded just to compare, the Ca~II H\&K region does NOT present background contamination.}
\label{fig:CaIIHKlines}
\end{figure}

\begin{figure}[t]
\centering
    \epsscale{0.75}
     \plotone
{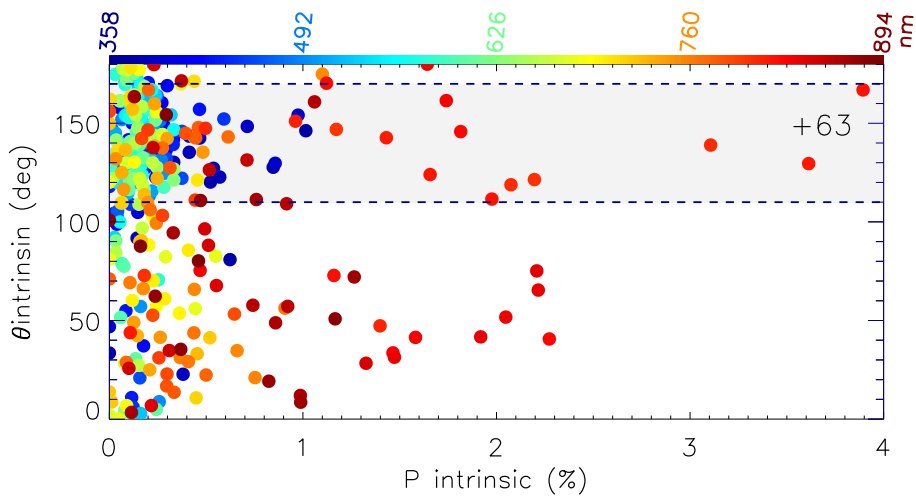}
\caption{Polarization versus polarization angle of SN2007sr (bin=15\AA{}). Strong polarization is distributed around two angles, 140$\arcdeg$ and 45$\arcdeg$. The first one colored in gray includes Ca~II, H\&K (blue dots) and the IR triplet (red dots). The 45$\arcdeg$ angle corresponds to sky polarization, which is strong and contaminates part of the CaII IR triplet profile (See Figure \ref{fig:skyThetavsPol}).}
\label{fig:ThetavsPol}
\end{figure}

\begin{figure}[t]
\centering
    \epsscale{0.75}
    \plotone
{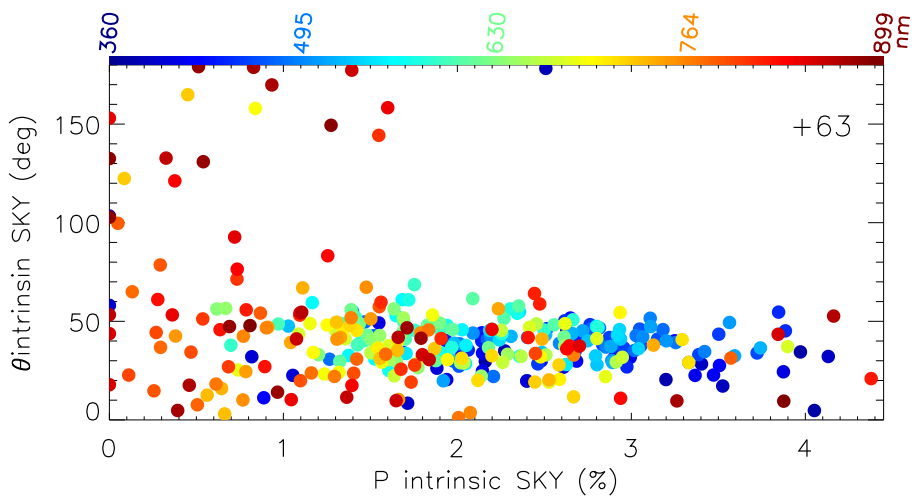}
\caption{Polarization versus polarization angle of the spectrum of the sky of SN 2007sr (bin=15\AA{}). Sky polarization is probably dominated by moonlight polarization, at the observation date the moon was at an angular distance of 140$\arcdeg$ away from the SN, at first quarter phase. The sky polarization is distributed along the 45$\arcdeg$ angle, in Figure \ref{fig:ThetavsPol} this corresponds to the lower angle zone, below the gray region.}
\label{fig:skyThetavsPol}
\end{figure}

\begin{figure}[t]
\centering
    \epsscale{0.75}
     \plotone
{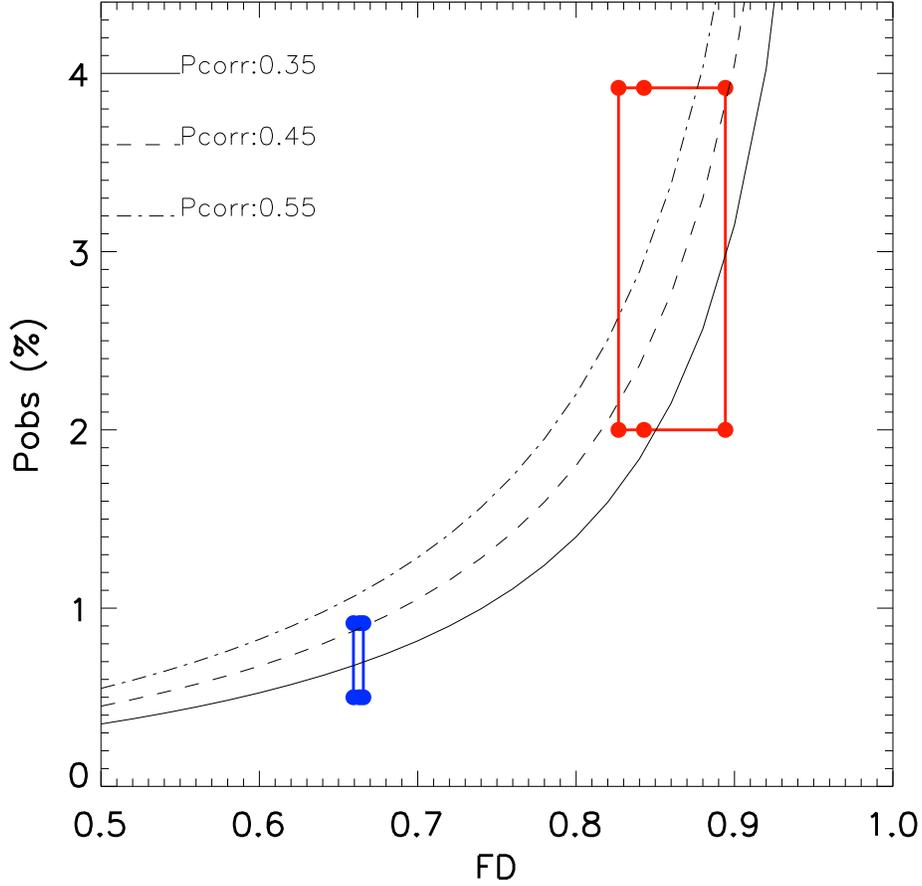} 
\caption{Observed line polarization of the Ca~II features as a function of the
fractional depth (FD) of the absorption line in the intensity spectrum,
based on the work of \citet{Tanaka2012ApJ...754...63T}.
Three estimates of FD for the Ca~II H\&K line (in blue) and Ca~II IR triplet
(in red) are given, corresponding to an upper, lower and average estimate of
the local continuum. 
The varying line polarization at each line is illustrated by the vertical lines,
between $2-4$\% for the infrared triplet and $0.5-1$\% for the H \& K lines.
The plotted lines correspond to equation (8) in
\citet{Tanaka2012ApJ...754...63T} for different values of the parameter they
call corrected polarization (P$_{\rm{corr}}$).
The model is very simple, and the uncertainties are large, but it is
nevertheless reassuring to find that the very different values of
observed polarization in the two calcium lines is consistent with
a single value of P$_{\rm{corr}}$, around 0.35-0.45.}
\label{fig:TanakaPcorr}
\end{figure}

\begin{figure}[t]
\centering
    \epsscale{1.1}
    \plottwo
%{/home/lavioleta/polariz/2007sr/figures/plotQUnewSN2007srBetab15.eps}{/home/lavioleta/polariz/2007sr/figures/plotQUnewSN2007srBetab30.eps}
{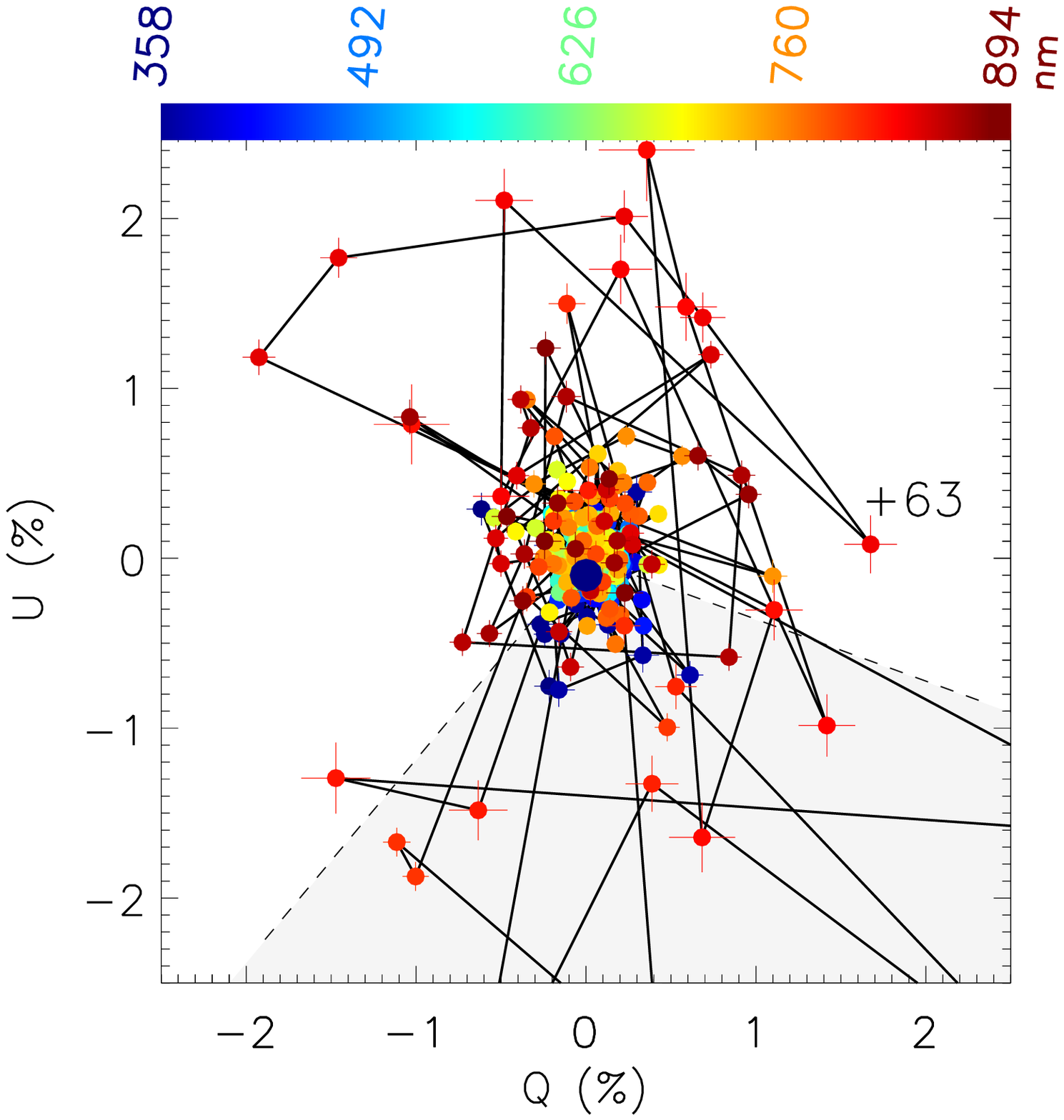}{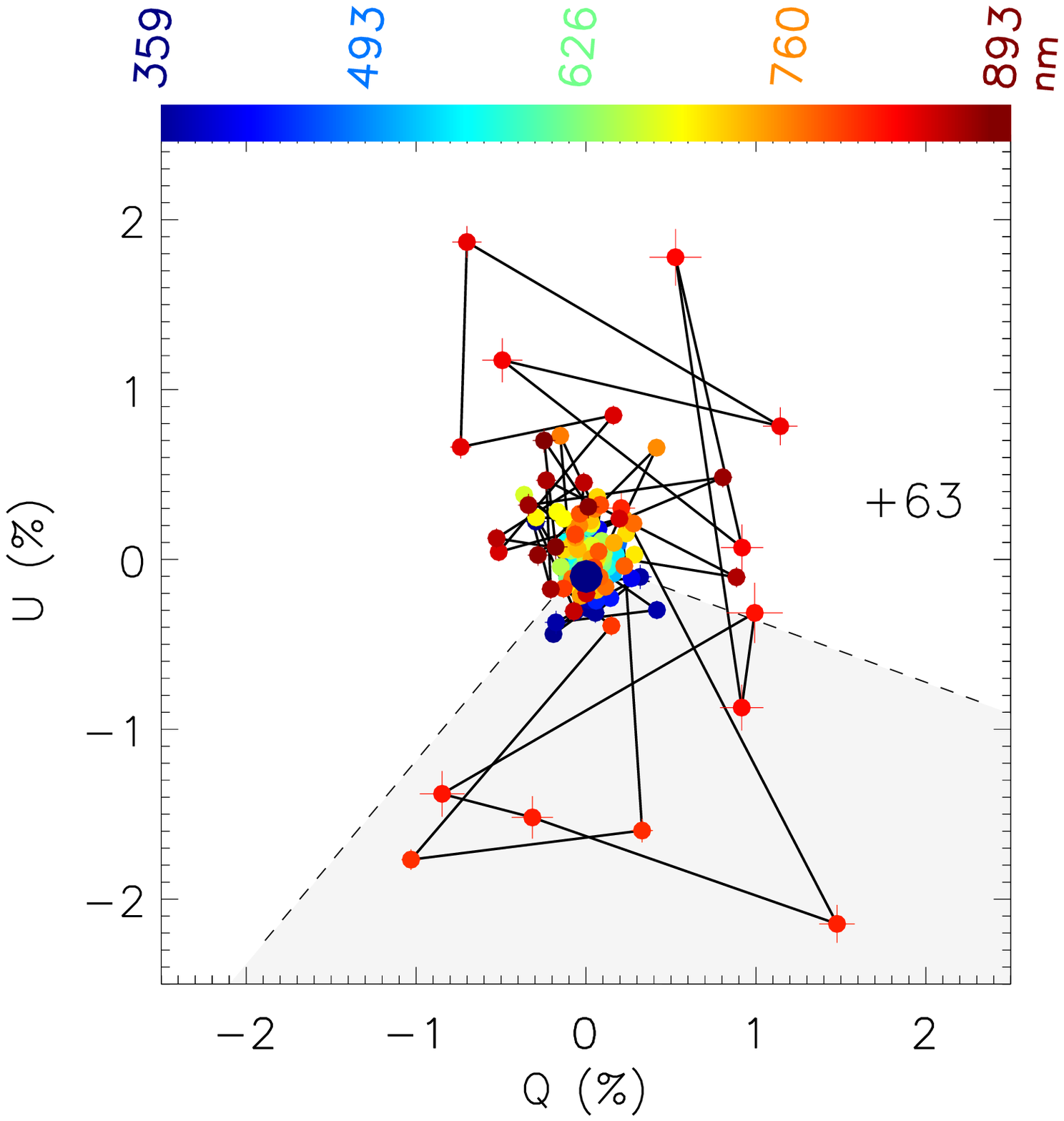}
\caption{Observed QU diagram of the entire wavelength range at +63 days after max, the left panel corresponds to bin=15\AA{} and the right panel to bin=30\AA{}. The Q$_{ISP}= 0$, U$_{ISP}=0.1$ and central wavelength W$_{ISP}= 550$ nm is marked by a black dot. Most of the polarization is concentrated around zero (null continuum polarization) except for a remarkable loop in the red portion of the spectrum, around Ca~II IR. The gray angular sector corresponds to the gray band shown in figure \ref{fig:ThetavsPol}. }
\label{fig:QUdiagram}
\end{figure}

\begin{figure}[t]
\centering
    \epsscale{1.1}
    \plottwo
%{/home/lavioleta/polariz/2007sr/figures/plotQUnewSN2007srCaIIIRb15.ps}{/home/lavioleta/polariz/2007sr/figures/plotQUnewSN2007srCaIIIRb30.ps}
{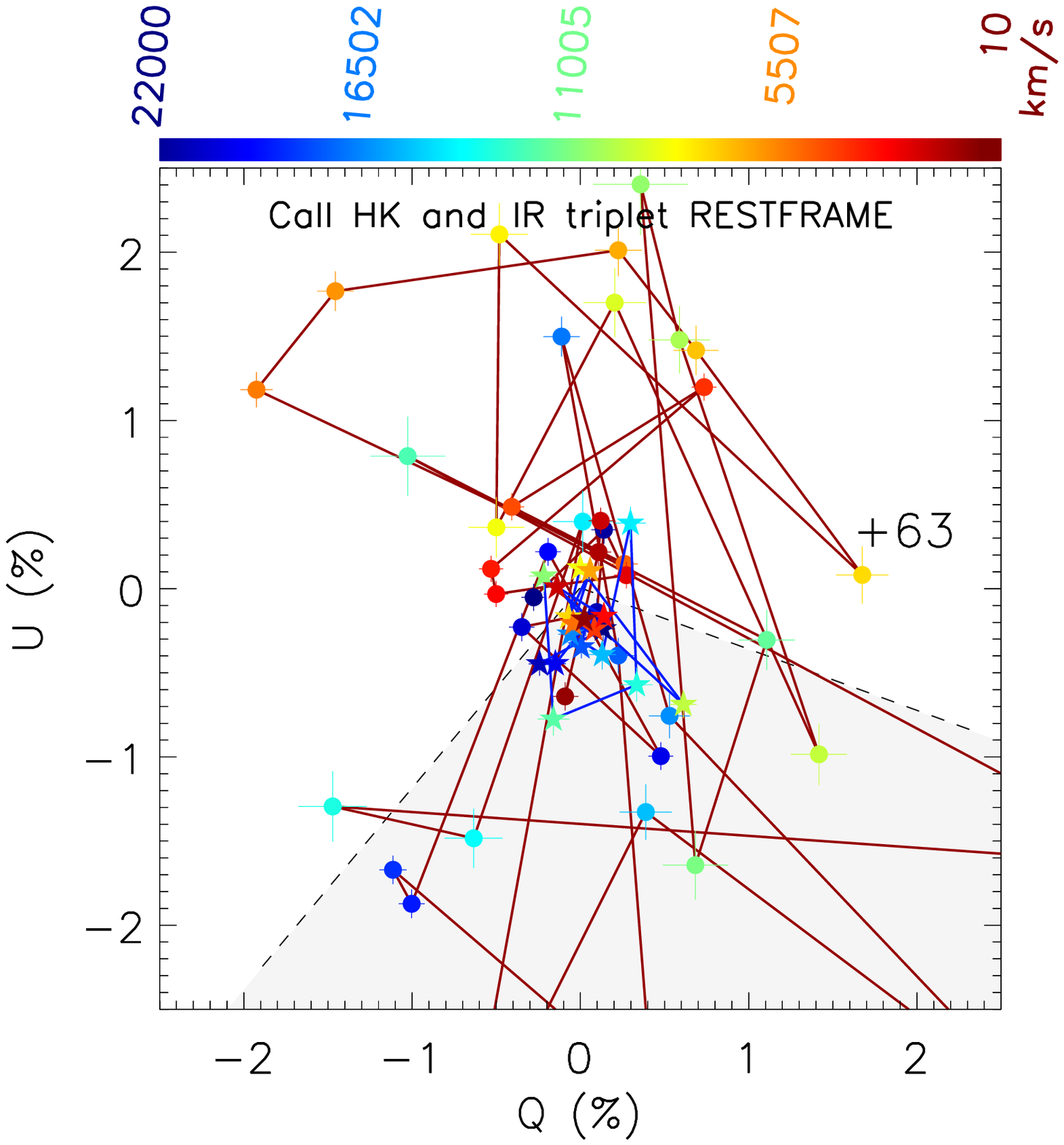}{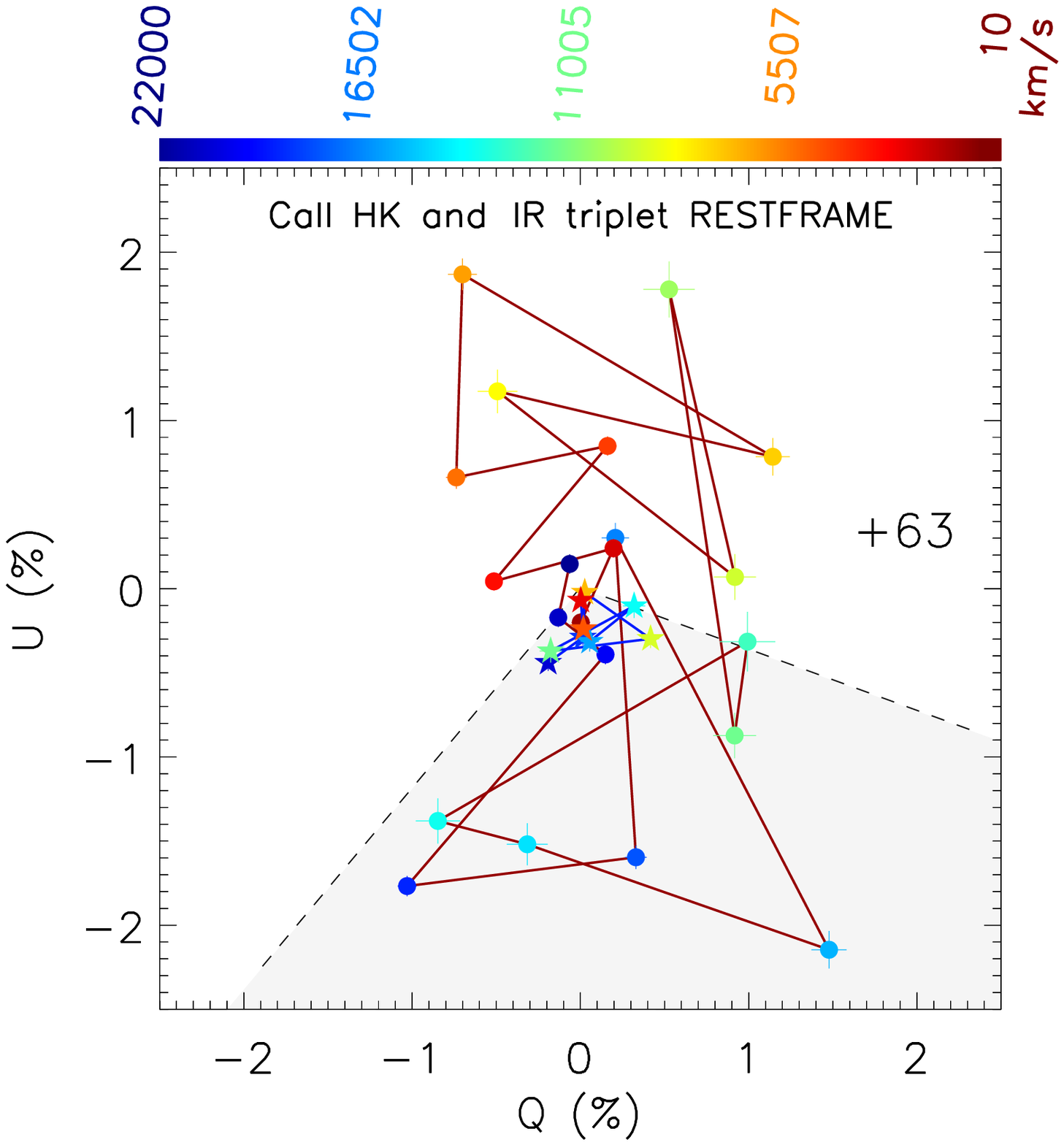}
\caption{Observed QU diagram of the Ca~II IR region, showing a loop of high velocities (left panel corresponds to bin=15\AA{} and the right panel to bin=30\AA{}). Blue and cyan dots, correspond to the velocity range 20000--13000 \kms, where the polarization measurements are uncontaminated by the sky, and they are also the highest polarization values. The gray angular sector corresponds to the gray band shown in figure \ref{fig:ThetavsPol}.}
\label{fig:QUdiagramCaIR}
\end{figure}

\citet{Umbriaco2007CBET.1174....1U} found high velocities in a spectrum of SN 2007sr taken closer to maximum light, and suggested that the SN matched the spectra of High Velocity Gradients (HVG) SNe \citep{Benetti2005ApJ...623.1011B}; further confirmed by \citet{Maeda2010Natur.466...82M} and \citet{Benetti2011}. % it is %
Consistent with this, our spectrum of SN~2007sr at 63 days after maximum light reveals high expansion velocities.
The most prominent absorption lines that exhibits fast velocities are those of Ca~II,  which span the range 5--18 $\times 10^{3}$ \kms.
High velocities at late phases, are consistent with the existence of high velocity components prior to maximum light, as it was the case for SN 2004dt and SN 2006X \citep{Wang2006ApJ...653..490W,Patat et2009A&A...508..229P}.
All the SNe used for comparison spectra in Figure \ref{fig:latespecs} (SN 1994D, 2003du and 2006X) showed high velocity components
in their early phases, which makes it tempting to deduce that SN2007sr too,  must have had High Velocity Features (HVF) in its early phases.
%
%SN 2006X registered expansion velocities of about 20500 \kms 10 days before maximum \citep{WangX2008ApJ...675..626W,Patat et2009A&A...508..229P} this was probably also the case of SN 2007sr, which could have showed this element at $\> 30000$\ \kms before maximum light (¿De donde sacaste este estimate?.

The very low polarization of the continuum adds one more case to consolidate the standard view that normal Type Ia SNe have only small, if any,
departures from global spherical symmetry; in particular towards later epochs
when we see deeper layers \citep{Filippenko_Leonard2004cetd.conf...30F,Leonard2005ApJ...632..450L,Wang_and_Wheeler2008ARA&A}.
The line polarization, however, shows a different picture.
The usual expectation is that line polarization, if present at early epochs, will decrease and disappear at later phases.
The best studied line is Si~II 6355. In most SNe observed around maximum light, and few days or weeks after maximum,
the polarization of the line peaks around maximum and then disappears with a time scale of weeks \citep{Wang2007Sci...315..212W}.
The same behaviour has been reported in polarization of Ca~II lines. SN~2001el showed a high velocity Ca II IR triplet, with a maximum line polarization $\sim$ 0.8\% at early times. This polarization decreased after maximum and was very small at a phase of 38 days \citep{Wang2003ApJ...591.1110W}.

It appears, however, that the Ca~II lines do not always follow this tendency.
SN 2007sr exhibits strong polarization in  Ca~II two months after maximum light, and it happens in a rapidly expanding feature.
\citet{Patat et2009A&A...508..229P} found high late time polarization of the Ca II IR triplet in SN 2006X, another rapidly expanding event observed from early phases until 39 days after maximum light.
They describe a re-polarization of the line and, based on the abundance distribution of a delayed detonation model of \citet{Hoflich2002ApJ...568..791H}, interpret it as a partial cover up of the photosphere in the line wings, where the Sobolev optical depth is smaller.
They also point out that the lack of significant line polarization at the inner Ca boundary ($v \sim$ 10000 \kms \,for the same model) is a
probable sign of mixing at that inner edge.
Due to the contamination by sky background, we cannot trace the polarization of the Ca~II IR triplet to these low velocities.
The Ca~II H\&K lines, however, which give a more certain estimate of the intrinsic
polarization, allows us to see the same effect in SN~2007sr.
The line polarization decrease sharply for velocities between 9000 and 8000 \kms,
which will be consistent with mixing up to a lower velocity.

Having just one spectrum of SN~2007sr, we do not know if the high polarization of Ca~II is a re-polarization at late times or if Ca~II was always polarized from maximum light onwards.
It is important to note that SN~2007sr adds to the case that {\em some} Type Ia SNe that are considered normal do show high line polarization at late phases.
We do not know why some do and some others do not show this late polarization.
From the data in hand however, it is tempting to connect this late time line polarization with the presence of HV Ca~II features. 
If this were the case the late time polarization could be due to interaction of the ejecta with circumstellar matter as suggested by \citet{Gerardy2004AAS...204.6308G}

Few supernovae have been observed with spectropolarimetry at phases as late as SN 2006X and 2007sr.
We already mentioned SN 2001el, another case 
is SN 2005ke, a subluminous event, which did not show Ca~II polarization to a statistically significant level either before maximum, or at $\sim$76 days after maximum \citep{Patat2012A&A...545A...7P}.
It is important to increase the database.
A larger, good quality dataset will allow us to connect this observable with others, like the
presence of high velocity features, and, eventually, help us to disentangle effects of the
explosion physics, from those of the aspect angle, or those of the interaction between the ejecta and circumstellar matter.

\acknowledgments
  P.Z., J.Q. \& A.C. acknowledge support by Iniciativa Cientifica Milenio (MINECON, Chile) through the
Millennium Center for Supernova Science (P10-064-F). 
P.Z. also acknowledges CONICYT, Chile (Beca de Doctorado).
The research of J.R.M. is supported through a Royal Society University Research Fellowship.
A.C. thanks for support through grants Basal CATA PFB 06/09 and FONDAP No. 15010003 from CONICYT, Chile.
P.A.H. was supported by the NSF grants 0703902 and 0708855.
%

% References
%

\newpage

\clearpage

% Las figuras van después de la bibliografía en los manuscritos (ver manual de estilo de la AAS)

\clearpage

\clearpage

\clearpage

%Theta vs Pol PLOT

%sky Theta vs Pol PLOT

\clearpage
%Pobs vs FD (Tanaka Pcorr)

%\clearpage

%QU PLOT

%QU PLOT


\begin{thebibliography}{}
%\parsep 0pt
%\itemsep -3pt

%ABC

\bibitem[Benetti et al.(2005)]{Benetti2005ApJ...623.1011B} Benetti, S., 
Cappellaro, E., Mazzali, P.~A., et al.\ 2005, \apj, 623, 1011 

%\bibitem[Benetti (2011)]{Benetti2011} Benetti, S.\ 2011, private communication 



\bibitem[Branch et al.(2005)]{Branch2005PASP..117..545B} Branch, D., Baron, E., Hall, N., Melakayil, M., \& Parrent, J.\ 2005, \pasp, 117, 545 

\bibitem[Branch et al.(1993)]{Branch1993AJ....106.2383B} Branch, D., Fisher, A., \& Nugent, P.\ 1993, \aj, 106, 2383 

\bibitem[Branch et al.(2008)]{Branch2008PASP..120..135B} Branch, D., Jeffery, D.~J., Parrent, J., et al.\ 2008, \pasp, 120, 135 

\bibitem[Burns et al.(2011)]{Burns2011AJ....141...19B} Burns, C.~R., Stritzinger, M., Phillips, M.~M., et al.\ 2011, \aj, 141, 19 


%DEF
\bibitem[Drake et al.(2007)]{Drake2007CBET.1172....1D} Drake, A.~J., Djorgovski, S.~G., Williams, R., et al.\ 2007, CBET, 1172, 1 

\bibitem[Fesen et al.(2007)]{Fesen2007ApJ...658..396F} Fesen, R.~A., 
H{\"o}flich, P.~A., Hamilton, A.~J.~S., et al.\ 2007, \apj, 658, 396

\bibitem[Filippenko 
\& Leonard(2004)]{Filippenko_Leonard2004cetd.conf...30F} Filippenko, A.~V., \& Leonard, D.~C.\ 2004, Cosmic explosions in three dimensions, 30

\bibitem[Folatelli et al.(2010)]{Folatelli2010AJ....139..120F} Folatelli, G., 
Phillips, M.~M., Burns, C.~R., et al.\ 2010, \aj, 139, 120 





%GHIJK
\bibitem[Gerardy et al.(2004)]{Gerardy2004AAS...204.6308G} Gerardy, C.~L., 
H{\"o}flich, P., Fesen, R.~A., et al.\ 2004, BAAS, 36, 777 

\bibitem[G{\'o}mez 
\& L{\'o}pez(1998)]{Gomez1998AJ....115.1096G} G{\'o}mez, G., \& L{\'o}pez, R.\ 1998, \aj, 115, 1096 


%BVRI Light Curves for 29 Type IA Supernovae
\bibitem[Hamuy et al.(1996)]{Hamuy1996AJ....112.2408H} Hamuy, M., Phillips, M.~M., Suntzeff, N.~B., et al.\ 1996, \aj, 112, 2408 




%%The Absolute Luminosities of the Calan/Tololo Type IA Supernovae
%\bibitem[Hamuy et al.(1996a)]{Hamuy1996AJ....112.2391H} Hamuy, M., Phillips, 
%M.~M., Suntzeff, N.~B., Schommer, R.~A., Maza, J., 
%\& Aviles, R.\ 1996, \aj, 112, 2391 


%%The Hubble Diagram of the Calan/Tololo Type IA Supernovae and the Value of HO
%\bibitem[Hamuy et al.(1996b)]{Hamuy1996AJ....112.2398H} Hamuy, M., Phillips, 
%M.~M., Suntzeff, N.~B., Schommer, R.~A., Maza, J., 
%\& Aviles, R.\ 1996, \aj, 112, 23


%\bibitem[Hamuy et al.(1996d)]{Hamuy1996AJ....112.2438H} Hamuy, M., Phillips, 
%M.~M., Suntzeff, N.~B., Schommer, R.~A., Maza, J., Smith, R.~C., Lira, P., 
%\& Aviles, R.\ 1996, \aj, 112, 2438 


\bibitem[Hamuy et al.(2000)]{Hamuy2000AJ....120.1479H} Hamuy, M., Trager, S.~C., 
Pinto, P.~A., Phillips, M.~M., Schommer, R.~A., Ivanov, V., 
\& Suntzeff, N.~B.\ 2000, \aj, 120, 1479 

\bibitem[Hatano et al.(1999)]{Hatano1999ApJ...525..881H} Hatano, K., Branch, D., 
Fisher, A., Baron, E., \& Filippenko, A.~V.\ 1999, \apj, 525, 881 
%High veloc ejecvta 1994D


\bibitem[Hillebrandt et al.(2000)]{Hillebrandt2000AIPC..522...53H} Hillebrandt, W., 
Niemeyer, J.~C., 
\& Reinecke, M.\ 2000, AIP Conf. Ser. 522, 53


\bibitem[H{\"o}flich(1991)]{Hoflich1991A&A...246..481H} H{\"o}flich, P.\ 1991, \aap, 246, 481 

\bibitem[H{\"o}flich et al.(2002)]{Hoflich2002ApJ...568..791H} H{\"o}flich, P., 
Gerardy, C.~L., Fesen, R.~A., \& Sakai, S.\ 2002, \apj, 568, 791 

\bibitem[H{\"o}flich 
\& Khokhlov(1996)]{Hoeflich_Khokhlov1996ApJ...457..500H} H{\"o}flich, P., \& Khokhlov, A.\ 1996, \apj, 457, 500 


\bibitem[Howell et al.(2001)]{Howell2001ApJ...556..302H} Howell, D.~A., 
H{\"o}flich, P., Wang, L., \& Wheeler, J.~C.\ 2001, \apj, 556, 302 


\bibitem[Ivanov et al.(2000)]{Ivanov2000ApJ...542..588I} Ivanov, V.~D., Hamuy, 
M., \& Pinto, P.~A.\ 2000, \apj, 542, 588 


\bibitem[Kasen et al.(2009)]{Kasen2009Natur.460..869K} Kasen, D., R{\"o}pke, F.~K., \& Woosley, S.~E.\ 2009, \nat, 460, 869 


\bibitem[Kasen et al.(2008)]{Kasen2008JPhCS.125a2007K} Kasen, D., Thomas, R.~C., R{\"o}pke, F., \& Woosley, S.~E.\ 2008, JPhCS, 125, 012007 

\bibitem[Khokhlov(1991)]{Khokhlov1991A&A...245L..25K} Khokhlov, A.~M.\ 1991, \aap, 245, L25 

%LMN
\bibitem[Lauberts 
\& Valentijn(1989)]{LaubertsandValentijn1989Msngr..56...31L} Lauberts, A., \& Valentijn, E.~A.\ 1989, The Messenger, 56, 31 

\bibitem[Leonard et al.(2005)]{Leonard2005ApJ...632..450L} Leonard, D.~C., Li, W., 
Filippenko, A.~V., Foley, R.~J., \& Chornock, R.\ 2005, \apj, 632, 450 


\bibitem[Liu et al.(1997)]{Liu1997ApJ...489L.141L} Liu, W., Jeffery, D.~J., 
Schultz, D.~R., et al.\ 1997, \apjl, 489, L141 
%CoII in late 1994D


\bibitem[Madore \& Steer (2007)]{Madore_and_Steer2007} Madore, B.~F., \& Steer, I.~P.\ 2007, 
Master List of Galxy Distances (Rev 2.2), http://ned.ipac.caltech.edu/level5/NED1D/intro.html


\bibitem[Maeda et al.(2010a)]{Maeda2010Natur.466...82M} Maeda, K., Benetti, S., Stritzinger, M., et al.\ 2010, \nat, 466, 82 

\bibitem[Maeda et al.(2010b)]{Maeda2010ApJ...708.1703M} Maeda, K., Taubenberger, S., Sollerman, J., et al.\ 2010, \apj, 708, 1703 


\bibitem[Maund et al.(2010a)]{Maund2010ApJ725L.167M} Maund, J.~R., H{\"o}flich, P., Patat, F., et al.\ 2010, \apjl, 725, L167 

\bibitem[Maund et al.(2010b)]{Maund2010ApJ...722.1162M} Maund, J.~R., Wheeler, J.~C., Wang, L., et al.\ 2010, \apj, 722, 1162 

%\bibitem[Mazzali et al.(2005)]{Mazzali2005ApJ...623L..37M} Mazzali, P.~A., Benetti, S., Altavilla, G., et al.\ 2005, \apjl, 623, L37 

%\bibitem[Naghizadeh-Khouei \& Clarke(1993)]{Naghizadeh-Khouei1993A&A...274..968N} Naghizadeh-Khouei, J., \& Clarke, D.\ 1993, \aap, 274, 968 


%\bibitem[Nugent et al.(1995)]{Nugent1995ApJ...455L.147N} Nugent, P., Phillips, M., Baron, E., Branch, D., \& Hauschildt, P.\ 1995, \apjl, 455, L147 

%OPQR
%\bibitem[Patat et al.(1996)]{Patat1996MNRAS.278..111P} Patat, F., Benetti, S., Cappellaro, E., et al.\ 1996, \mnras, 278, 111 


%2006X
\bibitem[Patat et al.(2009)]{Patat et2009A&A...508..229P} Patat, F., Baade, D., H{\"o}flich, P., et al.\ 2009, \aap, 508, 229 

\bibitem[Patat et al.(2012)]{Patat2012A&A...545A...7P} Patat, F., H{\"o}flich, P., Baade, D., et al.\ 2012, \aap, 545, A7 

\bibitem[Patat et al.(2011)]{Patat2011A&A...527A..91P} Patat, F., M{\"o}hler, S., O'Brien, K., et al.\ 2011, \aap, 527, A91 

\bibitem[Patat \& Romaniello (2006)]{PatatRomaniello2006PASP..118..146P} Patat, F., \& Romaniello, M.\ 2006, \pasp, 118, 146 

\bibitem[Phillips(1993)]{Phillips1993ApJ...413L.105P} Phillips, M.~M.\ 1993, \apjl, 413, L105

\bibitem[Pojmanski et al.(2008)]{Pojmanski2008CBET.1213....1P} Pojmanski, G., Prieto, J.~L., Stanek, K.~Z., \& Beacom, J.~F.\ 2008, CBET, 1213, 1 

%\bibitem[Perlmutter et al.(1997)]{Perlmutter1997ApJ...483..565P} Perlmutter, S., et 
%al.\ 1997, \apj, 483, 565 


\bibitem[Prieto et al.(2005)]{Prieto2005ASPC..339...69P} Prieto, J.~L., Rest, A., \& Suntzeff, N.~B.\ 2005, in ASP Conf. Ser, 339, Observing Dark Energy, 69 


%STUV
\bibitem[Saviane et al.(2008)]{Saviane2008ApJ...678..179S} Saviane, I., Momany, 
Y., da Costa, G.~S., Rich, R.~M., \& Hibbard, J.~E.\ 2008, \apj, 678, 179 

\bibitem[Schlegel et al.(1998)]{Schlegel1998ApJ...500..525S} Schlegel, D.~J., 
Finkbeiner, D.~P., \& Davis, M.\ 1998, \apj, 500, 525 

\bibitem[Schweizer et al.(2008)]{Schweizer2008AJ} Schweizer, F., et 
al.\ 2008, \aj, 136, 1482 

\bibitem[Serkowski et al.(1975)]{Serkowski1975ApJ...196..261S} Serkowski, K., 
Mathewson, D.~S., \& Ford, V.~L.\ 1975, \apj, 196, 261 

%\bibitem[Simmons \& Stewart(1985)]{Simmons1985A&A...142..100S} Simmons, J.~F.~L., \& Stewart, B.~G.\ 1985, \aap, 142, 100 


\bibitem[Stanishev et 
al.(2007)]{Stanishev2007A&A...469..645S} Stanishev, V., Goobar, A., Benetti, S., et al.\ 2007, \aap, 469, 645 
%2003du 480 days

%\bibitem[Stewart(1991)]{Stewart1991A&A...246..280S} Stewart, B.~G.\ 1991, \aap, 246, 280 

\bibitem[Tanaka et al.(2012)]{Tanaka2012ApJ...754...63T} Tanaka, M., Kawabata, 
K.~S., Hattori, T., et al.\ 2012, \apj, 754, 63 


\bibitem[Tonry et al.(2001)]{Tonry2001ApJ...546..681T} Tonry, J.~L., Dressler, 
A., Blakeslee, J.~P., et al.\ 2001, \apj, 546, 681 

\bibitem[Turatto et al.(2003)]{Turatto2003fthp.conf..200T} Turatto, M., Benetti, 
S., 
\& Cappellaro, E.\ 2003, in ESO Astrophys. Symp., From Twilight to Highlight: The Physics of Supernovae, 200 

\bibitem[Umbriaco et al.(2007)]{Umbriaco2007CBET.1174....1U} Umbriaco, G., 
Pietrogrande, T., di Mille, F., et al.\ 2007, CBET, 1174, 1 

%WXYZ

\bibitem[Wang et al.(2003)]{Wang2003ApJ...591.1110W} Wang, L., Baade, D., 
H{\"o}flich, P., et al.\ 2003, \apj, 591, 1110 

\bibitem[Wang et al.(2006)]{Wang2006ApJ...653..490W} Wang, L., Baade, D., H{\"o}flich, P., et al.\ 2006, \apj, 653, 490 

\bibitem[Wang et al.(2007)]{Wang2007Sci...315..212W} Wang, L., Baade, D., \& Patat, F.\ 2007, Science, 315, 212 
%\bibitem[Wang et al.(2008)]{WangX2008ApJ...675..626W} Wang, X., Li, W., Filippenko, A.~V., et al.\ 2008, \apj, 675, 626 


\bibitem[Wang \& Wheeler(2008)]{Wang_and_Wheeler2008ARA&A} Wang, L., \& Wheeler, J.~C.\ 2008, \araa, 46, 433

%\bibitem[Wang et al.(2008)]{WangX2008ApJ...675..626W} Wang, X., Li, W., Filippenko, A.~V., et al.\ 2008, \apj, 675, 626 

\bibitem[Yamanaka et al.(2009)]{Yamanaka2009PASJ...61..713Y} Yamanaka, M., Naito, H., Kinugasa, K., et al.\ 2009, \pasj, 61, 713 

                         
\end{thebibliography}
\end{document}